\documentclass[aps,prc,twocolumn,showpacs,superscriptaddress,groupedaddress]{revtex4}  
\usepackage{mathrsfs,amsmath,amssymb,amsthm}
\usepackage{amssymb,fmtcount}
\usepackage{graphicx,epsfig,latexsym,overpic,amssymb,color}
\usepackage{threeparttable}
\preprint{\today}

\begin{document}
\title{
Modeling survival probabilities of superheavy nuclei at high excitations
}
\author{C.Y. Qiao}
\affiliation{
State Key Laboratory of Nuclear Physics and Technology, School of Physics,
Peking University, Beijing 100871, China
}
\author{J.C. Pei}\email{peij@pku.edu.cn}
\affiliation{
State Key Laboratory of Nuclear Physics and Technology, School of Physics,
Peking University, Beijing 100871, China
}


\begin{abstract}
This work investigated the first-chance survival probabilities of highly excited compound superheavy nuclei in the prospect of synthesizing new superheavy elements.
The main feature of our modelings is the adoption of microscopic temperature dependent fission barriers in calculations of fission rates.
A simple derivation is demonstrated to elucidate the connection between Bohr-Wheeler statistical model and  imaginary free energy method, obtaining a new formula for fission rates.
The best modeling is chosen with respect to reproducing the experimental fission probability of $^{210}$Po.
Systematic studies of fission and  survival probabilities of No, Fl, Og, and $Z$=120 compound nuclei are performed.
Results show large discrepancies by different models for survival probabilities of  superheavy nuclei although
they are close for $^{210}$Po. We see that the first-chance survival probabilities of $Z$=120 are comparable to that of Fl and Og.
\end{abstract}
\maketitle
\section{introduction}

To synthesize the heaviest elements is one of the major science problems~\cite{Super2,Super4}.
To date, superheavy elements Z=107-113~\cite{Z107,Z108,Z109,Z110,Z111,Z112,Z113} have been synthesized in cold fusion reactions with the target Pb or Bi, and Z=114-118~\cite{Z114,Z115,Z116,Z117,Z118} have been synthesized in hot fusion reactions with the projectile $^{48}$Ca.
 The 7th row of the periodic table of elements has been completed.
 In order to produce superheavy elements in the 8th row, experimental attempts to synthesize Z=119 and 120 were performed in laboratories using  reactions such as $^{58}$Fe+$^{244}$Pu~\cite{FePu} at JINR,  $^{51}$V+$^{248}$Cm~\cite{VCm} at RIKEN,  and $^{64}$Ni+$^{238}$U, $^{50}$Ti+$^{249}$Bk, $^{50}$Ti+$^{249}$Cf, $^{54}$Cr+$^{248}$Cm~\cite{UNi,TiCf,CrCm} at GSI, but no evidence of new elements was observed.
  Currently, the measured limit of the residue cross section is less than 65 fb for next superheavy nuclei~\cite{TiCf}.
 The main issue is to design the optimal combination of beam-target nuclei and the bombarding energy.
 In this context, reliable theoretical guidance would be valuable for such extremely difficult experiments.

Theoretically, the synthesis process of superheavy nuclei can be described as the capture-fusion-evaporation reaction. In this procedure the  residue cross section is written as~\cite{Wsur}:
\begin{equation}
\sigma_{ER}=\sigma_{\rm cap}P_{\rm CN}W_{\rm sur}
\end{equation}
which depends on the capture cross-section $\sigma$$_{\rm cap}$,  the fusion probability \emph{P}$_{\rm CN}$ to compound nuclei, and the survival probability \emph{W}$_{\rm sur}$ of excited compound nuclei. The survival probabilities of compound nuclei are determined by the competition between the neutron emission rates and fission rates.
There are many models been developed to predict the synthesis of superheavy nuclei~\cite{NCM1,NCM2,FBD,twostep,DNS}.
The combined modeling of three steps can results in large uncertainties.
In particular, the surprising large cross sections of hot fusion reactions indicate that microscopic calculations of survival probabilities are essential~\cite{hamilton,peiPRL,zhuyi2017}.
Thus reliable modelings of survival probabilities by extrapolation is important since such experimental constraints in the superheavy region are rare.

Conventionally, the statistical models have been widely used for calculations of survival probabilities of compound nuclei~\cite{stat1,stat2,stat3}. The statistical models
can be traced back to Weisskopf's work for particle evaporations in 1937~\cite{evaporation} and Bohr-Wheeler's work for fission rates in 1939~\cite{Bohr-wheeler}.
The Bohr-Wheeler statistical model is based on classical transition state theory, which relies on fission barriers and level densities.
Actually, the fission barriers and level density parameters could be dependent on temperatures (or excitation energies), nuclear deformations, and shell structures.
Consequently, statistical models have to adopt parameterized energy dependent corrections to  fission barriers and level density parameters~\cite{prc105-014620}.
In the standard statistical model, the fission barrier is energy independent.  However, energy dependent fission barriers are necessary
to obtain reasonable fission observables  such as survival propabilities ~\cite{itks} and fission product yields~\cite{zhaojie2019}.

The thermal fission  rates can also be calculated by the dynamical Kramers model~\cite{Kramer} and the imaginary free energy approach (Im\emph{F})~\cite{IMF1}. Im\emph{F} can in principle describe the fission rates from low to high excitation energy in the quantum statistical framework.  The fission barrier heights and the curvatures around the equilibrium point and the saddle point are essential inputs for Kramers and Im\emph{F} methods~\cite{Zhuyi2016}. This can also be microscopically estimated but has rarely been discussed. Strutinsky pointed out a systematic difference between Karmers model and Bohr-Wheeler model~\cite{Strutinsky,Schmidt}. It is valuable to elucidate the connections between Bohr-Wheeler, Kramers, and Im\emph{F} models.

In this paper, our main goal is to study the survival probabilities of superheavy nuclei in the microscopic framework based on the finite-temperature Skyrme-Hartree-Fock+BCS approach~\cite{Goodman}.
The fission barriers are given in terms of free energies and are energy dependent. Then the fission rates are obtained with the imaginary free energy(Im\emph{F}) method~\cite{IMF1,Zhuyi2016}, the Bohr-Wheeler models~\cite{Bohr-wheeler} and the Kramers model~\cite{Kramer}. The neutron evaporation rates can be obtained by the standard statistical model.  For comparison, the neutron evaporation rates are also estimated by the density of neutron gas around nuclear surfaces~\cite{Zhu yi 2014}.
The connections between three fission models are discussed. To benchmark different models, the fission probability of $^{210}$Po up to high excitations are studied before being extrapolated to the superheavy region.

\section{THEORETICAL FRAMEWORK}
\subsection{Finite-Temperature Hartree-Fock+BCS Calculations}

In this work, the fission barriers are calculated with Skyrme-Hartree-Fock+BCS at finite temperatures (FT-BCS)~\cite{Goodman}.
Previously we have studied the fission barriers with the finite-temperature Hartree-Fock-Bogoliubov method~\cite{peiPRL,Pei2010},
which is computationally more expensive.
For systematic calculations, FT-BCS calculations are performed with SkyAX solver in cylindrical coordinate spaces~\cite{SkyAX}.
The Skyrme interaction SkM$^{*}$~\cite{SkM} and the mixed pairing interaction~\cite{pairing} are used. Note that SkM$^{*}$
is developed particularly for descriptions of fission barriers.


In FT-BCS, the normal density $\rho$ and the pairing density $\tilde{\rho}$ at a finite temperature are modified as~\cite{Goodman}:
\begin{equation}
\rho_{T}(r)=\sum_{i}[u_{i}^{2}f_{i}+\upsilon_{i}^{2}(1-f_{i})]|\phi_{i}(r)|^{2}
\end{equation}
\begin{equation}
 \tilde{\rho}_{T}(r)=\sum_{i} u_{i} \upsilon_{i}(1-2f_{i})|\phi_{i}(r)|^{2}
\end{equation}
where \emph{f}$_{i}$ = 1/(1+e$^{E_{i}/kT}$)(\emph{kT} is the temperature in MeV) is the temperature dependent factor; \emph{E}$_{i}$ is the quasiparticle energy; \emph{k} is the Boltzmann constant.
The entropy \emph{S} is evaluated as:
\begin{equation}
 S=-k\sum_{i}[f_{i}\ln f_{i}+(1-f_{i})\ln (1-f_{i})],
\end{equation}
The temperature dependent fission barriers are calculated in terms of the free energy \emph{F}=\emph{$E_T$}$-$\emph{TS} , where \emph{$E_T$} is the intrinsic binding energy based on the temperature-dependent densities.

For realistic calculations, we also need mass parameters which is obtained by the Cranking formula~\cite{Baran94}.
At a finite temperature, the resulted mass parameters have significant fluctuations as a function of deformations~\cite{Zhuyi2016}.
However, the mass parameters extracted from microscopic dynamical calculations are not much dependent on excitation energies~\cite{tanimura,qiangyu}.
Therefore we adopt the mass parameters at zero temperature in all calculations.

It has been demonstrated that uniform neutron-gas density distributions can be obtained in coordinate-space FT-HFB calculations~\cite{peiPRL,Zhu yi 2014}.
This also appears in coordinate-space FT-BCS calculations. With the uniform neutron gas density \emph{n}$_{gas}$, the neutron emission width $\Gamma_{n}$ is given by the nucleosynthesis formula~\cite{gammn-FT}:
\begin{equation}
\Gamma_{n}=\hbar n_{gas}\langle \sigma \upsilon \rangle
\label{ngas}
\end{equation}
where $\sigma$ is the neutron capture cross section and estimated by the geometric area $\pi$\emph{R}$^{2}$. $\langle$$\upsilon$$\rangle$ is the average velocity of the external gas. This method doesn't involve level densities, see details in Ref.~\cite{Zhu yi 2014}.

\subsection{Imaginary Free Energy Method}

The imaginary free energy method (Im\emph{F}) is in principle can describe a system's metastablility from quantum tunneling at low temperatures to  statistical decays at  high temperatures in a consistent framework~\cite{IMF1}.
In Im\emph{F}, the fission barrier is naturally given by temperature dependent free energies.
This method has been previously used to evaluate fission rates of compound nuclei~\cite{Zhuyi2016}.
At low temperatures, the Im\emph{F} formula for the fission width from excited systems is given as~\cite{IMF1}:
\begin{equation}\label{Gammf-IMF}
\begin{aligned}
\Gamma_{f}&=\frac{1}{Z_{0}} \frac{1}{2\pi \hbar} \int_{0}^{V_{b}}P(E) \exp(-\beta E) \, dE, \\
Z_{0}&=[2 \sinh(\frac{1}{2}\beta \hbar \omega_{0})]^{-1}
\end{aligned}
\end{equation}
where $\omega$$_{0}$ is the curvature or frequency around the equilibrium point at the potential valley; \emph{V}$_{b}$ is the barrier height; \emph{Z}$_{0}$ is the partition function; $\beta$ denotes $1/kT$; \emph{P}(E) is the barrier transmission probability.

For the fission probability at the high temperatures, the contribution is dominated by reflections above the barriers. In this case, the transmission probability \emph{P}(E) can be estimated by:
\begin{equation}\label{tran}
P(E)=\displaystyle\{1+\exp[2\pi(E^{}-V_{b})/ \hbar \omega_b]\} ^{-1}
\end{equation}
Then the fission width at high temperatures can be written as~\cite{IMF1}:
\begin{equation}
\Gamma_{f}=\frac{\omega_{b}}{2\pi} \frac{\sinh(\frac{1}{2}\beta \hbar \omega_{0})}{\sin(\frac{1}{2}\beta \hbar \omega_{b})}\exp(-\beta V_{b})
\end{equation}
where  $\omega$$_{b}$ is the  curvature at the saddle point of the barrier.
The curvatures $\omega$$_{0}$ and $\omega$$_{b}$   can be easily calculated with the
microscopic temperature dependent fission barriers and the mass parameters, as discussed in Ref.\cite{Zhuyi2016}.
There is a narrow transition in Im\emph{F} formulas from low to high temperatures, which is dependent on the critical temperature $\hbar\omega_b/2\pi$~\cite{IMF1}.
The Im\emph{F} method has been widely applied in chemistry reactions in a thermal bath. For nuclear fission studies,  Im\emph{F} shows that the fission lifetime
decreases very rapidly at low excitations and decrease slowly at high excitations~\cite{Zhuyi2016}. With temperature dependent fission barriers, it is a success for Im\emph{F}
to reveal that  the compound nucleus $^{278}$Cn in cold fusion can not survive at high excitations while $^{292}$Fl in hot fusion still has a considerable survival probability at high excitations~\cite{Zhuyi2016}.

\subsection{Bohr-Wheeler Statistical Model}
The Bohr-Wheeler statistical model has widely been used to calculate the survival probabilities of superheavy nuclei~\cite{stat1,stat2}.
This is also known as the transition state theory and is based on the micro-canonical statistics.
The width of neutron evaporation is given by~\cite{evaporation}:
\begin{equation}
\Gamma_{n}(E)=\frac{2m R^{2}}{\pi \hbar^{2} \rho_0(E)}\int_{0}^{E-S_{n}} \varepsilon_{n}\rho_0(E-S_{n}-\varepsilon_{n}) \, d\varepsilon_{n}
\label{stat-n}
\end{equation}
Here, \emph{m} is the neutron mass; \emph{R} is the radius of the compound nucleus; \emph{S}$_{n}$ is the neutron separation energy and $\rho_0$(\emph{E}) is the level density
at the equilibrium deformation.

The fission width can be calculated with the Bohr-Wheeler formula as~\cite{Bohr-wheeler}:
\begin{equation}\label{Gammf-stat}
\Gamma_{f}(E)=\frac{1}{2\pi \rho_0(E)}\int_{0}^{E-V_{b}} \rho_{s}(E-V_{b}-\varepsilon_{f}) T_{f}(\varepsilon_{f})\, d\varepsilon_{f}
\end{equation}
where  $\rho$$_{s}$(\emph{E}) is the level density at the saddle point, and \emph{T}$_{f}$($\varepsilon$$_{f}$) is the barrier transmission probability:
\begin{equation}\label{tran-pro}
T_{f}(\varepsilon_{f})=\{1+\exp[- \frac{2\pi \varepsilon_{f}}{\hbar \omega_{sd}}]\}^{-1}
\end{equation}
Usually $\hbar$$\omega$$_{sd}$ is taken as 2.2 MeV in calculations as mentioned in Ref.~\cite{stat3}.

The level density is calculated with the Fermi-gas model as ~\cite{Fermi model} :
\begin{equation}
\rho(E)=\frac{\sqrt{\pi}\exp(2\sqrt{a E})}{12 a^{1/4} E^{5/4}}
\label{leveld}
\end{equation}
where \emph{a} is the level density parameter taken as the usual \emph{a}=\emph{A}/12 MeV, the level density parameter at the saddle point is taken as \emph{a}$_{sd}$=1.1\emph{a}.
Note that there are modified formulas of energy dependent level densities~\cite{prc105-014620},
and associated parameters are dependent on the nuclear region.

\subsection{Connection between Statistical Model and Im\emph{F}}

It is interesting to study the connections between statistical model and Im\emph{F} for fission rates and survival probabilities of compound nuclei.
By Combining Eqs.(\ref{Gammf-IMF}) and (\ref{tran}), the fission width is given as:
\begin{equation}\label{Gammf-IMF2}
\begin{aligned}
\Gamma_{f}^{ImF}&=Z_{0}^{-1} \int_{}^{} \frac{1}{2\pi \hbar} \frac{1}{1+\exp(-\frac{2\pi (E-V_{b})}{\hbar \omega})}\exp(-\beta E)\,dE \\
Z_{0}&=[2\sinh(\frac{1}{2}\beta \hbar \omega_{0})]^{-1}
\end{aligned}
\end{equation}
where \emph{V}$_{b}$ is the fission barrier height.
If we use $\sinh$(x)=$\frac{e^{x}-e^{-x}}{2}$ and define \emph{x}=\emph{E}-\emph{V}$_{b}$,
then Eq.(\ref{Gammf-IMF2}) can be written as:
\begin{equation}\label{Gammf-IMF3}
\Gamma_{f}^{ImF}=\frac{\omega_{0}\hbar}{2\pi\hbar T} \int_{}^{} \frac{1}{1+\exp(-\frac{2\pi x}{\hbar\omega})} \exp(-\frac{x}{T})  \exp(-\frac{V_{b}}{T}) \, dx
\end{equation}

For the level density in statistical models,  $\ln$$\rho$$_{}$(E) is a smooth curve in terms of \emph{E}. When $\Delta E$ is small, we have:
\begin{equation}
\ln \rho_{}(E+\Delta E)=\ln\rho_{}(E)+[\frac{d \ln\rho_{}(E)}{dE}]\Delta E
\end{equation}
Since $\frac{1}{T_{}}$=[$\frac{d \ln\rho{}(E)}{dE}$] is defined by the evaporation model, so we have:
\begin{equation}\label{rho}
 \rho_{}(E+\Delta E)=\rho_{}(E)\exp(\frac{\Delta E}{T_{}})
\end{equation}
where \emph{T}$_{}$ is the nuclear temperature.
By combining Eqs.(\ref{Gammf-stat}), (\ref{tran-pro}) and (\ref{rho}), we can get:
\begin{equation}\label{Gammf-stat2}
 \Gamma_{f}^{BW}=\frac{1}{2\pi \rho_{0}(E)} \int_{0}^{E-V_{b}} \frac{\rho_{s}(E-V_{b})}{1+\exp(-\frac{2\pi k}{\hbar \omega})} \exp(-\frac{k}{T}) \, dk
\end{equation}
By comparing Eqs.(\ref{Gammf-IMF3}) and (\ref{Gammf-stat2}), the connection between  two methods is:
\begin{equation}
 \Gamma_{f}^{BW}= \frac{T}{\omega_{0}} \frac{\rho_{s}(E-V_{b})}{\rho_{0}(E)} \exp(\frac{V_{b}}{T}) \Gamma_{f}^{ImF}
 \label{connec}
\end{equation}

Since the two methods have different advantages and disadvantages, we now derive a new formula to
calculate the fission width,
\begin{equation}
 \Gamma_{f1}= \frac{\rho_{s}(E-V_b)}{\rho_{0}(E)} \frac{ \omega_{b}T}{2\pi \omega_{0}} \frac{\sinh(\frac{1}{2}\beta \hbar \omega_{0})}{\sin(\frac{1}{2}\beta \hbar \omega_{b})}
 \label{gf1}
\end{equation}
This formula is close to the Bohr-Wheeler model.
In this case, the calculations include the temperature dependent fission barrier heights in $\rho_s$ and also the
influence of the micro-canonical statistics. This formula doesn't need to do integral in statistical calculations of fission rates. It has been pointed out there is
a difference between Bohr-Wheeler fission model and the Kramers model by a factor $\omega_0/T$~\cite{Schmidt},
then we have another new formula for fission rates as,
\begin{equation}
\begin{aligned}
 \Gamma_{f2}&=\frac{\omega_{0}}{T}\Gamma_{f}^{BW} \\
     &= \frac{\rho_{s}(E-V_b)}{\rho_{0}(E)} \frac{ \omega_{b}}{2\pi} \frac{\sinh(\frac{1}{2}\beta \hbar \omega_{0})}{\sin(\frac{1}{2}\beta \hbar \omega_{b})}
\end{aligned}
 \label{gf2}
\end{equation}
This formula includes $\omega_0/T$, being consistent with Im\emph{F} and Kramers model at high temperatures~\cite{Kramer}.
At very high excitations, the barrier height $V_b$ is small compared to the excitation energy $E$.
By using the relation in Eq.(\ref{rho}) and Eq.(\ref{leveld}), then Eq.(\ref{connec}) becomes:
 \begin{equation}
 \Gamma_{f}^{BW}=\frac{T}{\omega_{0}} \frac{\exp[2\sqrt{E}(\sqrt{a_{sd}}-\sqrt{a})]}{(a_{sd}/a)^{1/4}} \Gamma_{f}^{ImF}
  \label{gf3}
\end{equation}
If we assume the level density parameters $a_{sd}$=$a$ at the limit of extremely high excitations, that means quantum shell effects are completely lost,
then we have,
 \begin{equation}
 \Gamma_{f}^{BW}=\frac{T}{\omega_{0}} \Gamma_{f}^{ImF}
\end{equation}
This demonstrates that results of Bohr-Wheeler model in micro-canonical ensemble  are close to
that of Im\emph{F} in heat bath at extremely high temperatures, except for the
prefactor $T/\omega_0$.   Note that Im\emph{F} is also close to the Kramers model at high temperatures~\cite{IMF1}.

\section{Results and discussions}

\begin{figure}[htbp]
\centering
\includegraphics[width=0.45\textwidth]{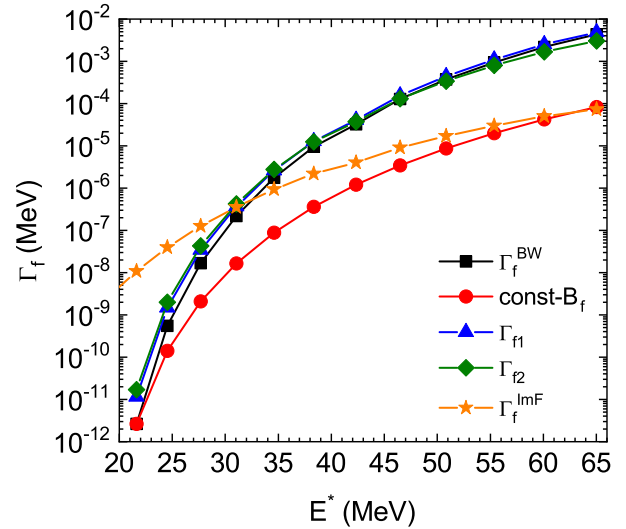}
\caption{
(Color online)
Calculated fission widths  $\Gamma$$_{f}$ of $^{210}$Po as a function of excitation energies E$^{*}$ with different models.
The energy dependent fission barriers are used except for const-B$_f$ within Bohr-Wheeler model.
$\Gamma_f^{\rm BW}$, $\Gamma_{f1}$, $\Gamma_{f2}$, $\Gamma_f^{\rm ImF}$ are obtained with Eqs.(\ref{Gammf-stat}), (\ref{gf1}), (\ref{gf2}), (\ref{Gammf-IMF}), respectively.
                                                              \label{FIG1}
}
\end{figure}

Firstly, we calculate the fission probabilities of $^{210}$Po
to benchmark various models. $^{210}$Po has accurate experimental energy dependent fission probabilities~\cite{210Po1,210Po2}.
Fig.\ref{FIG1} shows the calculated fission widths with different modelings.
In these calculations, the energy dependent fission barriers are adopted, which are taken from microscopic
finite-temperature Hartree-Fock+BCS calculations.
We can see that the fission widths calculated with Bohr-Wheeler model, the formula $\Gamma_{f1}$ (Eq.\ref{gf1}), the formula $\Gamma_{f2}$ (Eq.\ref{gf2})
are close. This verified the correctness of Eqs.(\ref{gf1}) and (\ref{gf2}). In this case, the role of the factor $\omega_0/T$ is not significant.
The results calculated by Im$F$ is also shown, which is very different from the Bohr-Wheeler results.
For example, the ratio difference estimated by Eq.({\ref{gf3}}) is 26.3 at 65 MeV if we assume $a_{sd}=1.1a$.
This explained that the significant differences between Bohr-Wheeler model and Im$F$ is due to the different level density parameters between saddle point and equilibrium point.
Note that in the original Bohr-Wheeler statistical model, the fission barrier is energy independent.
We also did calculations with a constant fission barrier height of 19.59 MeV based on Bohr-Wheeler model.
We can see that the fission widths with constant fission barriers are much smaller at high excitations.
This demonstrated that the essential role of energy dependent fission barriers in calculations of fission rates.

\begin{figure}[htbp]
\centering
\includegraphics[width=0.49\textwidth]{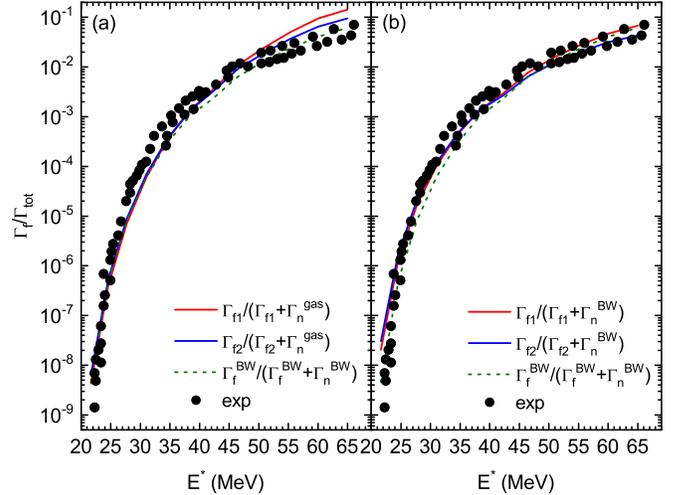}
\caption{
(Color online)
The results of $\Gamma$$_{f}$/$\Gamma$$_{tot}$ as a function of E$^{*}$ for $^{210}$Po.
Experimental data is taken from Ref.~\cite{210Po1,210Po2}.
Different fission widths and neutron emission widths are adopted. See text for details.                                                                   \label{FIG2}
}
\end{figure}

\begin{figure}[htbp]
\centering
\includegraphics[width=0.49\textwidth]{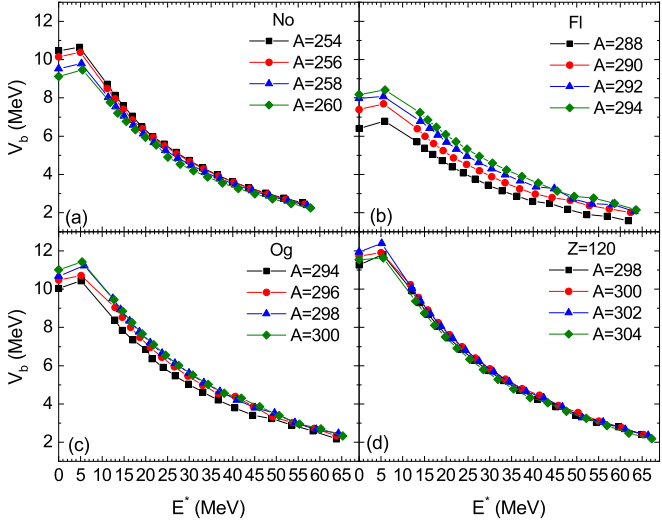}
\caption{
(Color online)
Calculated fission barrier heights of No (a), Fl (b), Og (c), Z=120(d) isotopes are shown as a
function of excitation energies.
                                                                   \label{FIG3}
}
\end{figure}

To compare with the experimental fission probabilities $\Gamma$$_{f}$/$\Gamma$$_{tot}$ of  $^{210}$Po,
the neutron evaporation widths have to be calculated.
The neutron evaporation widths can be calculated by the standard statistical model (Eq.\ref{stat-n}) or by
the microscopic neutron gas model (Eq.\ref{ngas}). In both calculations, the same nuclear radius $R$ is used for
the neutron-reaction cross section $\pi R^2$. The calculated fission probabilities are shown in Fig.\ref{FIG2}.
The experimental $\Gamma$$_{f}$/$\Gamma$$_{tot}$ of  $^{210}$Po are taken from proton and $\alpha$ induced fission reactions~\cite{210Po1,210Po2}.
It can be seen that all the calculations can well reproduce the experimental data.
The neutron gas model leads to a slightly larger fission probability or smaller survival probability.
The factor $\omega_0/T$ included in $\Gamma$$_{f2}$ results in slightly reduced fission probabilities at high excitations.
Other calculations with Im$F$ or constant fission barriers can not reproduce the experimental data and are not shown.

\begin{figure}[htbp]
\centering
\includegraphics[width=0.49\textwidth]{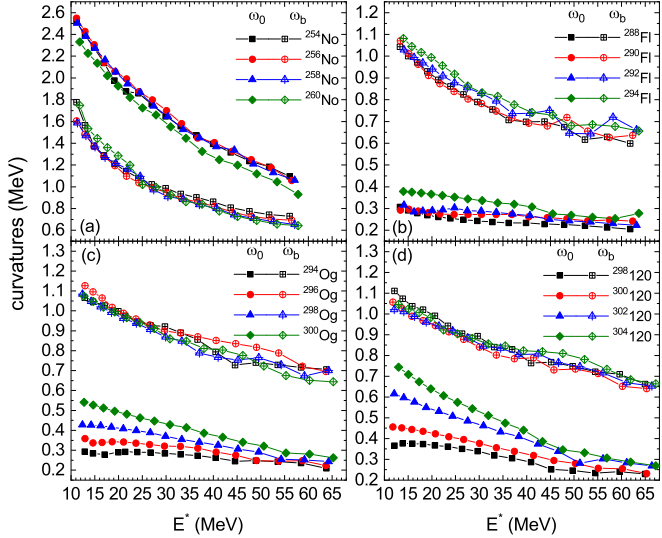}
\caption{
(Color online)
Calculated curvatures around the equilibrium point($\omega$$_{0}$) and the barrier saddle point($\omega$$_{b}$) as a function of excitation energy,
for No (a), Fl (b), Og (c) and Z=120 (d) isotopes.
                                                               \label{FIG4}
}
\end{figure}

Next we study the fission and survival probabilities of superheavy compound nuclei with the modelings that can reproduce the fission probabilities of $^{210}$Po.
In this work, four elements No, Fl, Og and Z=120 are selected for study.
The fission barrier heights as a function of excitation energy E$^{*}$ are shown in Fig.\ref{FIG3}.
We see that Z=120 isotopes have considerable fission barriers even at high excitations.
For $^{298-304}$120 isotopes, the barrier heights are not sensitive to the neutron numbers.
On the other hand, the barrier heights are much dependent on neutron numbers for $^{288-294}$Fl isotopes.
For Fl and Og isotopes, more neutrons result in enhanced fission barriers.
Note that the microscopic energy dependence of fission barriers of superheavy nuclei can be very different from empirical models.

In addition to fission barrier heights, the curvatures of fission barriers also play an important role in calculations of fission rates.
Fig.\ref{FIG4} displays the calculated energy dependent curvatures $\omega_0$ and $\omega_b$, corresponding to the equilibrium point and the saddle point, respectively.
 Generally, $\omega_0$ and $\omega_b$ decrease with increasing excitation energies, which lead to reduced fission rates.
For No isotopes, $\omega_0$ is larger than $\omega_b$.
For Fl, Og and Z=120 isotopes, $\omega_0$ is very small, that means the fission valley is soft.
Actually the ground state deformations of these nuclei are slightly oblate.
The very small $\omega_0$ can greatly enhance the stabilities of these nuclei against fission at high excitations.

\begin{figure*}[htbp]
\centering
\includegraphics[width=0.9\textwidth]{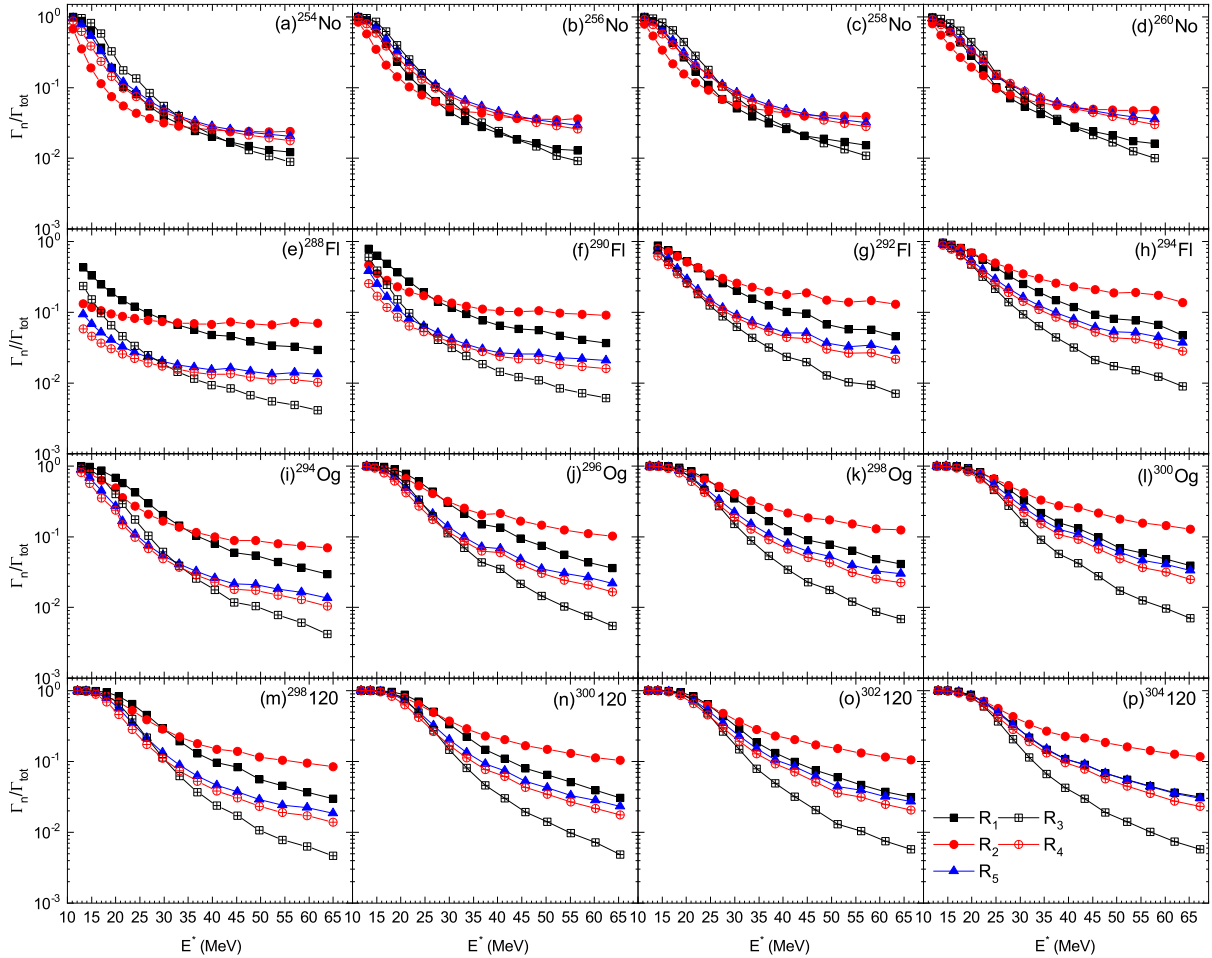}
\caption{
(Color online)
Calculated first-chance survival probabilities $\Gamma$$_{f}$/$\Gamma$$_{tot}$ of No, Fl, Og and Z=120 isotopes with
various modelings. Results of $^{254,256,258,260}$No are shown from (a) to (d); $^{288,290,292,294}$Fl are shown from (e) to (d);
$^{294,296,298,300}$Og are shown from (i) to (l); $^{298,300,302,304}$120 are shown from (m) to (p).
Different modelings:
 $R_1=\frac{\Gamma_n^{gas}}{\Gamma_n^{gas}+\Gamma_{f2}}$, $R_2=\frac{\Gamma_n^{BW}}{\Gamma_n^{BW}+\Gamma_{f2}}$,
$R_3=\frac{\Gamma_n^{gas}}{\Gamma_n^{gas}+\Gamma_{f1}}$, $R_4=\frac{\Gamma_n^{BW}}{\Gamma_n^{BW}+\Gamma_{f1}}$, and $R_5=\frac{\Gamma_n^{BW}}{\Gamma_n^{BW}+\Gamma_f^{BW}}$,
are shown for comparison.
                                                                  \label{FIG5}
}
\end{figure*}

Finally the first-chance survival probabilities $\Gamma$$_{n}$/$\Gamma$$_{tot}$  of superheavy nuclei are calculated, as shown in Fig.\ref{FIG5}.
Different modelings for the survival probabilities: $R_1=\frac{\Gamma_n^{gas}}{\Gamma_n^{gas}+\Gamma_{f2}}$, $R_2=\frac{\Gamma_n^{BW}}{\Gamma_n^{BW}+\Gamma_{f2}}$,
$R_3=\frac{\Gamma_n^{gas}}{\Gamma_n^{gas}+\Gamma_{f1}}$, $R_4=\frac{\Gamma_n^{BW}}{\Gamma_n^{BW}+\Gamma_{f1}}$, and $R_5=\frac{\Gamma_n^{BW}}{\Gamma_n^{BW}+\Gamma_f^{BW}}$,
are adopted for comparison. These modelings all are good for descriptions of fission probabilities of $^{210}$Po.

 In Fig.\ref{FIG5} (a)-(d), $\Gamma$$_{n}$/$\Gamma$$_{tot}$ results for $^{254,256,258,260}$No are shown.
  For $^{254}$No, results from different approaches are close. The discrepancies between different modelings increase with increasing neutron numbers.
 We see that $R_1$ and $R_3$ are close, and  $R_2$ and $R_4$ are close. This means that
 for No isotopes, the role of $\omega_0/T$ is not significant. At high excitations, $R_1$ (and $R_2$) is slightly larger than $R_3$ (and $R_4$)
 due to reduced $\omega_0/T$. It is on the contrary at low excitations.
 With the same fission width, the survival probabilities with $\Gamma_n^{gas}$ from the neutron gas model is obviously smaller than that with
 the statistical model at high excitations.
At low excitations, the survival probabilities with $\Gamma_n^{gas}$ are larger.

Fig.\ref{FIG5} (e)-(h) displays the survival probabilities of $^{288,290,292,294}$Fl isotopes.
For $^{288}$Fl, the fission barriers are lower than others, as shown in Fig.\ref{FIG3}.
The survival probabilities are much smaller than 1, and discrepancies are large at low excitations.
For $^{292,294}$Fl with higher fission barriers, the first-chance survival probabilities all are close to 1 at low excitations.
In experiments, the compound nuclei $^{290,292}$Fl are produced in hot fusion of $^{48}$Ca+$^{242,244}$Pu~\cite{Z114}.
Generally the survival probabilities increases with increasing neutron numbers for Fl isotopes.
Different from No isotopes, we see that with the same neutron evaporation width, the survival probabilities with $\Gamma_{f2}$
are considerably larger than that with $\Gamma_{f1}$.
This is because $\omega_0/T$ is very small for Fl isotopes at high excitations, which can significantly reduce fission widths.
With the same fission widths, the survival probabilities with $\Gamma_n^{BW}$ is much larger than that with $\Gamma_n^{gas}$
at high excitations.
For Fl isotopes,  $R_4$ and $R_5$ are in the middle between different modelings, while they are among the largest for No isotopes.

The survival probabilities of $^{294,296,298,300}$Og isotopes are shown in Fig.\ref{FIG5}(i)-(l).
For Z=120 isotopes, the results are shown in Fig.5(m)-(p).
Note that the compound nucleus $^{297}$Og is produced in hot fusion of $^{48}$Ca+$^{249}$Cf~\cite{Z118}.
The pattern of calculated survival probabilities for Og and 120 isotopes are very similar to that of $^{292,294}$Fl.
The combination of $\Gamma_n^{BW}$ and $\Gamma_{f2}$ leads to the largest survival probabilities.
Indeed, the factor $\omega_0/T$ is very small for Fl, Og and 120 isotopes at high excitations.
This can be traced back to the dynamical Kramers model and usually been ignored in statistical calculations.
Note that the Bohr-Wheeler model results in a much larger fission width about 1$\thicksim$2 MeV at high excitations in the superheavy region.
Actually fission becomes very dissipative at high excitations~\cite{qiangyu}.
If the fission width is about 1$\thicksim$2 MeV by the statistical model, then the later part of the fission from saddle to scission is not negligible at high excitations in real-time fission dynamics
~\cite{qiangyu2}.
The statistical model also results in larger neutron widths than the neutron gas model.
We speculate that the level density model used in both fission and neutron evaporation can have some offset effects.
In all cases, $R_4$ is close to $R_5$, which verified the correctness of the new formula.
The compound nuclei for synthesizing element 120 are $^{299,302}$120 in attempted  experiments~\cite{FePu,UNi,TiCf,CrCm}.
It can be seen that the first-chance for the survival probability of 120 is comparable to Fl and Og isotopes.

Current calculations still invoke the phenomenological level densities, while employ the microscopic energy dependent fission barriers.
This is essential for the micro-canonical ensemble.
Our results demonstrated that it is necessary to take into account the different level density parameters between the equilibrium and saddle point.
Previously the deformation and energy dependent level density parameters can be obtained by calculating  $S/2T$, or $E/T^2$, or $S^2/4E$~\cite{zhuyi2017}.
However, with these level densities,  the fission probabilities of $^{210}$Po can not be well reproduced.
With the current level densities and statistical model, the fission widths of superheavy nuclei would be as large as
1$\thicksim$2 MeV at high excitations, which is questionable.
The reliable calculations of energy dependent level densities is another challenge~\cite{future}.
The most optimistic estimations $R_2$ are given by the $\Gamma_{f2}$ and $\Gamma_n^{BW}$.
The neutron gas model always results in the lower limit of survival probabilities, which is also shown in $^{210}$Po.
For realistic calculations to guide experiments, the survival probabilities after multiple neutron emissions have to be studied in the future.

\section{summary}

In summary, we studied the first-chance survival probabilities of heavy and superheavy nuclei at high excitations with microscopic temperature
dependent fission barriers. There have been several experimental attempts to synthesis new superheavy elements Z=119 and 120.
Thus it is of interests to investigate various theoretical modelings of fission rates and survival probabilities.
With a simple derivation, we demonstrated the relation between the Bohr-Wheeler statistical model and the imaginary free energy method.
The derivation results in a new formula for fission rates.
To verify different modelings, the fission probabilities of $^{210}$Po have been reproduced, demonstrating the essential role of energy dependent fission barriers.
Next we studied the survival probabilities of No, Fl, Og and Z=120 isotopes with these modelings, which have large discrepancies in the superheavy region.
We see that the curvatures $\omega_0$ of the fission valley are very small for selected Fl, Og and 120 nuclei, which can
greatly enhance their survival probabilities.
On the other hand, the microscopic neutron gas model results in smaller neutron emission widths and reduced survival probabilities.
Generally, the first-chance survival probabilities of Z=120 nuclei at high excitations are comparable to that of Fl and Og nuclei.
In the future, for more realistic modelings, survival probabilities after multiple neutron emissions will be studied.

\acknowledgments
We thank the useful discussions with G. Adamian, A. Nasirov and F.R. Xu.
 This work was supported by  National Key R$\&$D Program of China (Contract No. 2018YFA0404403),
 and the National Natural Science Foundation of China under Grants No.11975032, 11790325, 11835001, and 11961141003.

\end{document}